**This is the accepted manuscript (postprint) of the following article:**
N. Abbasnezhad, N. Zirak, M. Shirinbayan, S. Kouidri, E. Salahinejad, A. Tcharkhtchi, F. Bakir, *Controlled release from polyurethane films: Drug release mechanisms*, Journal of Applied Polymer Science, 138 (2021) 50083.
https://doi.org/10.1002/app.50083

# Controlled release from polyurethane films: drug release mechanisms

N. Abbasnezhad [1, 2*], N. Zirak [1, 3], M. Shirinbayan [1, 2], S. Kouidri [4], E. Salahinejad [3], A. Tcharkhtchi [2], F. Bakir [1]

[1]*Arts et Metiers Institute of Technology, CNAM, LIFSE, HESAM University, 75013 Paris, France*

[2]*Arts et Metiers Institute of Technology, CNAM, PIMM, HESAM University, 75013 Paris, France*

[3] *Faculty of Materials Science and Engineering, K. N. Toosi University of Technology, Tehran, Iran*

[4] *Sorbonne University, UPMC Univ, Paris 6 UFR 919, 75005 Paris, France*

Emails: nader.zirak@ensam.eu, mohammadali.shirinbayan@ensam.eu, smaine.kouidri@limsi.fr, salahinejad@kntu.ac.ir, abbas.tcharkhtchi@ensam.eu, farid.bakir@ensam.eu

**Abstract**

In this study, polyurethane-films loaded diclofenac were used to analyze the drug release kinetics and mechanisms. For this purpose, the experimental procedures were developed under static and dynamic conditions with different initial drug loads of 10, 20, and 30%. In the dynamic condition, to better simulate the biological flow, drug release measurements were investigated at flow rates of 7.5 and 23.5 ml/s. These values indicate the flow rate of the internal carotid artery (ICA) for a normal state of a body and for a person during the exercise, respectively. The experimental data were analyzed and adjusted by Higuchi, Korsmeyer-Peppas, First-order, zero-order, and Peppas-Sahlin models in order to understand the mechanisms contributed. Finally, Drug release mechanisms were specified by investigating the model correlation coefficients. Experimental results showed that increasing the flow rate and initial drug loads enhance drug liberation. In addition, the rate of release is more influenced by the drug dosage in the static state. The analysis revealed that diffusion, burst, and osmotic pressure are the principal mechanisms contributed. Moreover, Fickian type was the dominant mechanism at all duration of release. However, it was discovered using Peppas-Sahlin model that the contribution of the diffusion mechanism decreases with increasing flow rate and initial dosage. Furthermore, the tests at different drug dosages showed that the number of stages in medication release profile is independent of the flow rate and the medicine percentage. One can conclude that the drug release kinetic in static state is more influenced by drug dosage compare to dynamic state.

*Corresponding author: navideh.abbasnezhad@ensam.eu

## 1. Introduction

The application and improvement of drug delivery systems in order to improve the safety and efficacy of conventional drugs have been the focus of many studies. Due to the importance of achieving appropriate drug release profiles from carriers as well as having properties such as biocompatibility, biodegradability, or non-biodegradability and mechanical properties has led to the widespread use of polymers [1, 2].

Drug delivery systems (DDS) in the form of the polymeric matrix have drawn interest in the area of therapeutic. Hydrophilic polymers are in this choice because they are generated with pore spaces which facilitate the release of the drug, especially as for low aqueous soluble drugs . In this regard, chemical and physical properties of the polymer such as glass transition temperature, permeability, viscosity, degradability, and concentration of polymers are the influential criteria that control the drug release from the polymer carrier [3]. Thus, all the characteristics referring to the polymer and drug, take account in the drug release profile [4]. Drug release refers to the definition where the drug solutes/non-solutes migrate from the initial position of the carrier to the outer side of the carrier to contact with the medium and consequently release in the medium [4]. Studies focused on achieving the systems with reducing toxicity, side effects, and enhancing the effectiveness of treatment-controlled release. Controlled drug release is affected by some parameters to improve the therapy's efficacy. These include, for example, the released time rate, global time, and position accuracy [5].

Drug delivery systems (DDS) in the form of the polymeric matrix have drawn interest in the area of therapeutic. Hydrophilic polymers are in this choice because they are generated with pore spaces which facilitate the release of the drug, especially as regards for low water soluble drugs. The time that the polymer is contacted with the aqueous medium diverse phenomena can occur like, degradation, dissolution, swelling, and etc. [6]. This can result in different drug release profiles by the mechanisms of diffusion, convection, burst, ion exchange, osmotic pressure and, etc. Depending on the necessities of the therapy, each of these mechanisms can take importance to the others [7].

The effects of the presence of each mechanism can differ depending on the polymer properties. As an example, the water swollen, and non-swollen polymers have been compared [8, 9]. It was revealed that the diffusion of the drug in the water swollen polymer was increased compared to the non-swollen polymer due to the creation of the free volume fraction. However, it is not always the same, as another study indicated that swelling of the non-degradable polymer decreases the release rate. This is for the reason that, it increases the path of the drug in the matrix for attaining to the release medium [8, 9].

Polyurethane has always been used in biomedical applications due to the good biocompatibility and mechanical properties [4, 10-15]. Polyurethane due to show the favorable ability to drug release profiles in various modes such as microparticles, films, and also as a coating on implants and stents has been the focus of many studies. In some studies, it is noted that the release from the non-degradable polymeric system is controlled by diffusion. In this case, the parameters which are taken importance are the thickness and hydrophilicity of the polymeric matrix; on the other hand, the solubility of the drug charged in the polymer also takes value. However, in a study by Huynh et al. [16] they have studied the release behavior of the chlorhexidine diacetate from the non-degradable PU and found that the release was followed by zero order mechanism. In the case of diffusion-controlled mechanisms, the reservoir and matrix type of the DDS is in the main categories. Indeed, matrix type is more favorable as it accepts not the risk of depleting the drug by the tearing of the cover layer. Furthermore, it utilizes a straightforward process of fabrication [17-19].

The other parameters which are important in the release are related to the properties of the drug. One of the types of the drugs used in the case of cardiovascular stenting is the anti-inflammatory drugs. Diclofenac is one of the most prescribed nonsteroidal anti-inflammatory drugs which inhibits the pain and reduces the inflammation [20-22]. Where different indicators such as the solubility, particle size, hydrophilicity of drug, and etc. will affect the release profile [23-25]. The drug particles in the matrix can have two kinds: they can dissolve in the polymer solvent therefore covalent bonds with the polymer are made or they can be dispersed in the matrix in the particle shape. The first type of degradation of the polymer affects the drug release data. However, for the second case degradation rate of the polymer, drug loading, solute size, swelling, and solubility influence the release rate [9].

Consequently, according to the polymer and the drug is chosen, various mechanisms of release can contribute to the release of the drug from the DDS. These mechanisms can continue the dissolution of the drug and DDS, diffusion of the liquid into the DDS and diffusion of the drug from DDS to the release medium, swelling, degradation, and erosion of the DDS, osmotic pressure and etc. [26]. Depending on the portion of participation, each of these mechanisms can represent the dominated mechanism of release or can be negligible. Also, depending on the rate of each mechanism occurring during the time of the release, it can be rate controlling on the whole period of the release [27].

In this study, the results related to the experiments conducted with charged polyurethane (PU) have been demonstrated. The experiments have been performed with varying two parameters: 1) flow rates of 0, 7.5 and 23.5 ml/s where there is a lack of study on it and, 2) drug percentages of 10%, 20%, and 30%wt. Then the steps related to the whole release were identified and the mechanisms have been analyzed. Subsequently, the contribution of the identified mechanisms has been calculated. The flow rate and drug ratio caused a significant effect on the mechanism of drug release from the carrier.

## 2. Experimental procedure

### *2.1. Materials*

#### 2.1.1. Polyurethane (PU)

In this analysis the substance used is non-degradable polyurethane which is the result of the synthesis of the hardener (isocyanate type 4,4-diphenylmethylene diisocyanate (MDI)) with resin Gyrothane 639 which is composed of polyol with $I_{OH}$ value of 336 (mg KOH/g), dye, and catalyst. The substance prepared from the hardener-resin combination in a proportion of 2:5 according to the datasheet, casted in a mold and heated at 50 °C for about 30 min. The related items were supplied from the RAIGI company.

#### 2.1.2. Diclofenac

Diclofenac as a non-steroidal anti-inflammatory drug is the agent loaded in PU samples for drug release tests [14]. It was purchased from Genevrier Laboratory in granular shape with a density of approximately 450.7 mg / ml. The solubility of the drug at a temperature of 37 °C in water is about 5.554 g/L. The particle size was determined by SEM images: around 40 μm ≤ particle size ≤ 160 μm.

### *2.2. Preparation of carriers and drug loading*

For drug release experiments, polymer films with the dimensions of 30×5×2 $mm^3$ were prepared. This dimension is considered an enlarged dimension of the strut of a stent. A process of molding and heating in an oven at the temperature of the 50°C for about 30 minutes was performed. For the samples with the active substance in the initial preparation of the mixture, a certain dosage of the drug (the mass ratio of drug/(drug + polymer): 10%, 20% and 30%wt) was added to the resin. After homogeneously dispersing the drug in the polyol the hardener was added to the mix. Finally, the prepared mix was poured in the mold and then it was put in the oven. Drug particles were maintained in the form of the granules after mixing. Figure 1 shows the schematic of this procedure. Scanning Electronic Microscope

"HITACHI 4800 SEM" has been used to analyze the morphology of the microstructure of the specimen throughout the study.

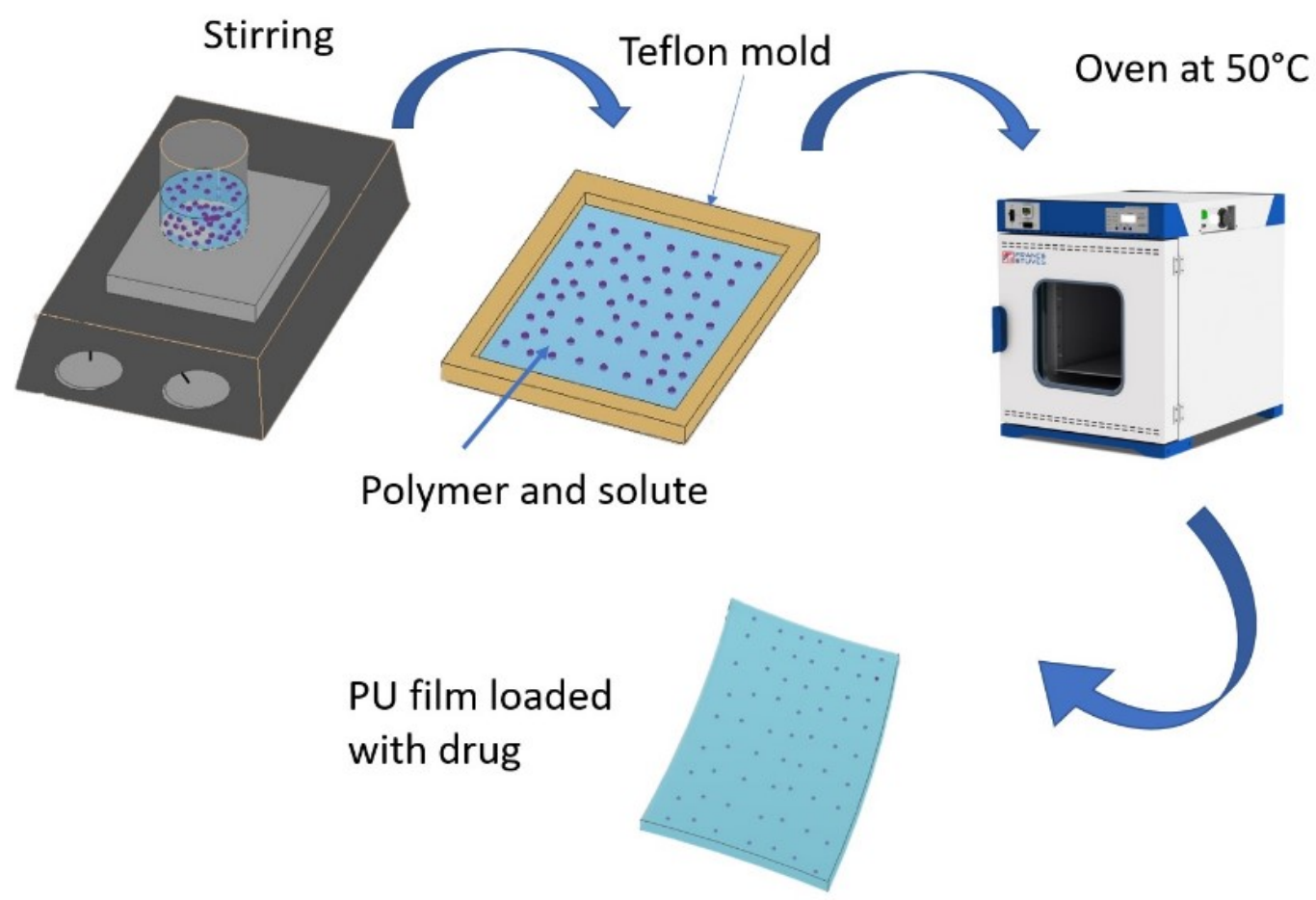

**Figure 1.** Process of casting for preparing PU films with the diclofenac dispersed inside

### *2.3. Test bench components*

The test facility used in this experiment is represented in figure 2. This equipment was designed to perform the test from static flow to the different flow rates by the aid of a pump and inverter. Valves are for adjusting the flow rate. A tank with the capacity of 10 liters is employed, although to avoid the risk of saturation of liquid, each two days of test it was refreshed with new one.

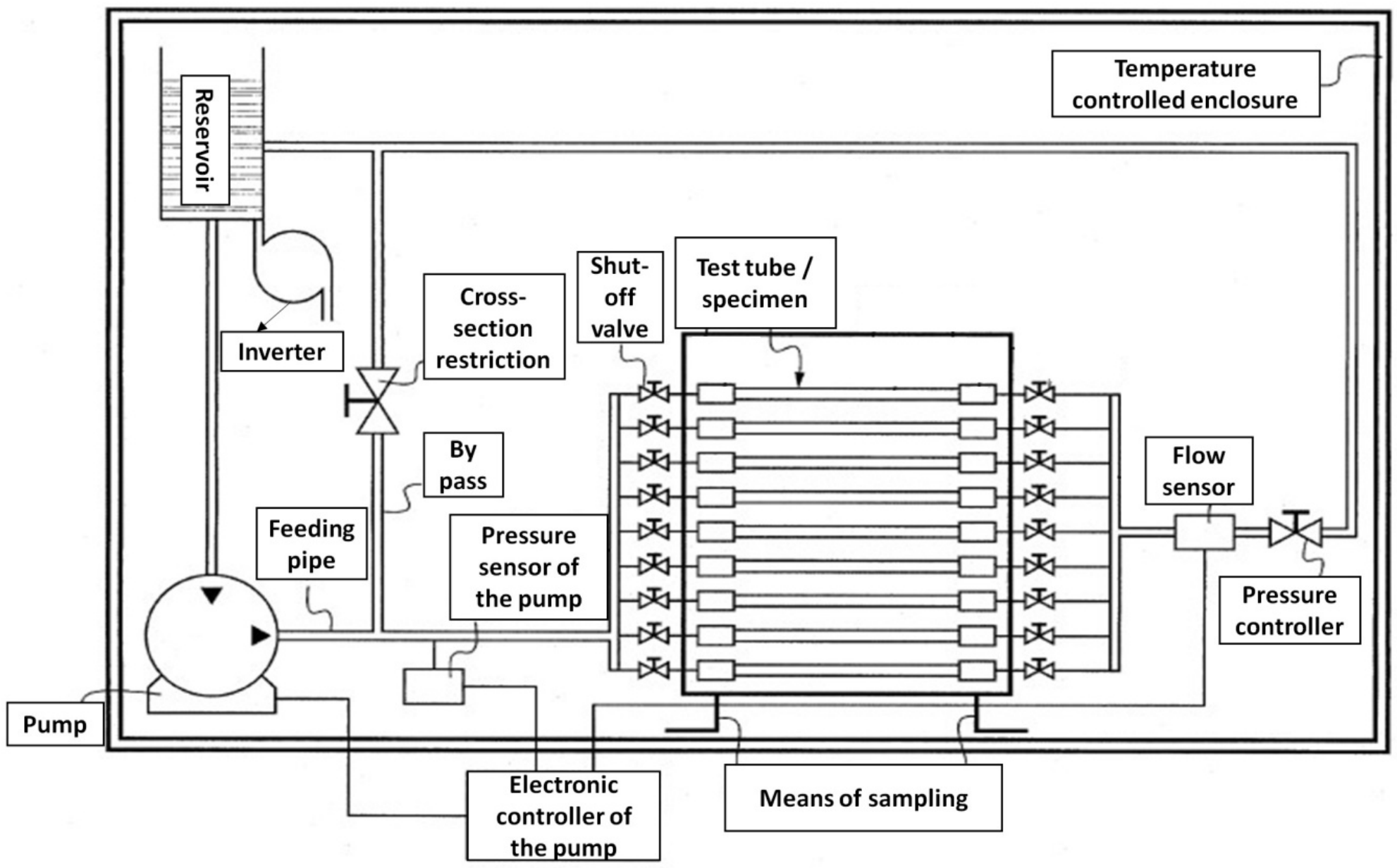

**Figure 2.** Equipment elements for tests at different flow rates

## 3. Experimental results and discussions

### *3.1. Drug release procedure and associated measurements*

Measurement methods were performed according to the following method. In the first step, the samples were dried in the oven at 50°C for one hour to remove any moisture. The samples were weighed immediately after cooling in the desiccator ($m_0$). The second step is to put the samples in the in-vitro test. Further the samples are collected from the test after a specified period for each test, to begin the following stage. The third step comprises cleaning the surface of the taken samples with a dry cloth to remove any water present. These samples then weighed ($m_1$). After that, to remove all the water in the film, the samples were transferred to an oven at 50 °C, after ensuring that all the water absorbed by the film was removed (after the sample weight was stabilized), the film was weighed ($m_2$). The equations for measuring the water absorption and drug release are listed below:

$$\text{Water absorption (\%)} = \frac{m_1 - m_2}{m_0} \times 100 \qquad \text{Eq. 1}$$

$$\text{Drug release (\%)} = \frac{m_0 - m_2}{\text{initial mass of drug}} \times 100 \qquad \text{Eq. 2}$$

where $m_0$ represents the sample mass at the initial state; $m_1$ the mass after wiping $m_2$ the mass after drying in the oven.

### *3.2. Water absorption at diverse flow rates and drug percentages*

Figure 3 (a) shows the water absorption of the PU loaded with diverse drug percentages of 0%, 10%, 20%, and 30% at the static state. It is noted that the water penetrates to the PU.0. Q: 0 (PU with zero percentage of the drug at the flow rate of zero ml/s) up to 5%, where it is augmented to the 70% for the samples with 30% of the drug after 10 days of immersion.

Therefore, the presence of the hydrophilic drug, carries out a significant role in water absorption. It is remarkable that the values of the percentage of water absorption are affected as well by the variation of the density of water and diclofenac which are 997 and 450.7 mg/l, respectively.

Figure 3 (b) shows the results of the water absorption for three various percentages of the drug at the flow rate of 7.5 ml/s; the concept is repeating once more for the effect of the drug percentage on water absorption in the flow condition. This figure indicates that the water absorption is increased by increasing the drug percentage.

By comparing the three curves of the figure 3(a), (b), and (c), one can note that the equilibrium water absorption for the samples with the same percentage of drug (for example 10%) at different flow rates gives rather the same value (~30%). Therefore, the equilibration value of water absorption is independent of the flow rate and it firmly depends on the composition of the sample (the percentage of the drug loaded). Further, it will affect the release behavior, whereas all the mechanisms mentioned attributing to the release are highly related to the water/liquid absorption. In the case of the more drug percentage due to the more quantity and larger diameter of the pores, water absorption increases where the free volume in the matrix can handle more solvent molecules. This process goes on until reaching an equilibrium value. Applying the equation of Korsemeyer-Peppas for the water absorption results gives the n value lower than 0.5. Therefore it indicates that water absorption and hence swelling in all the experimental tests are controlled by the diffusion mechanism (Fickian diffusion) [28-30].

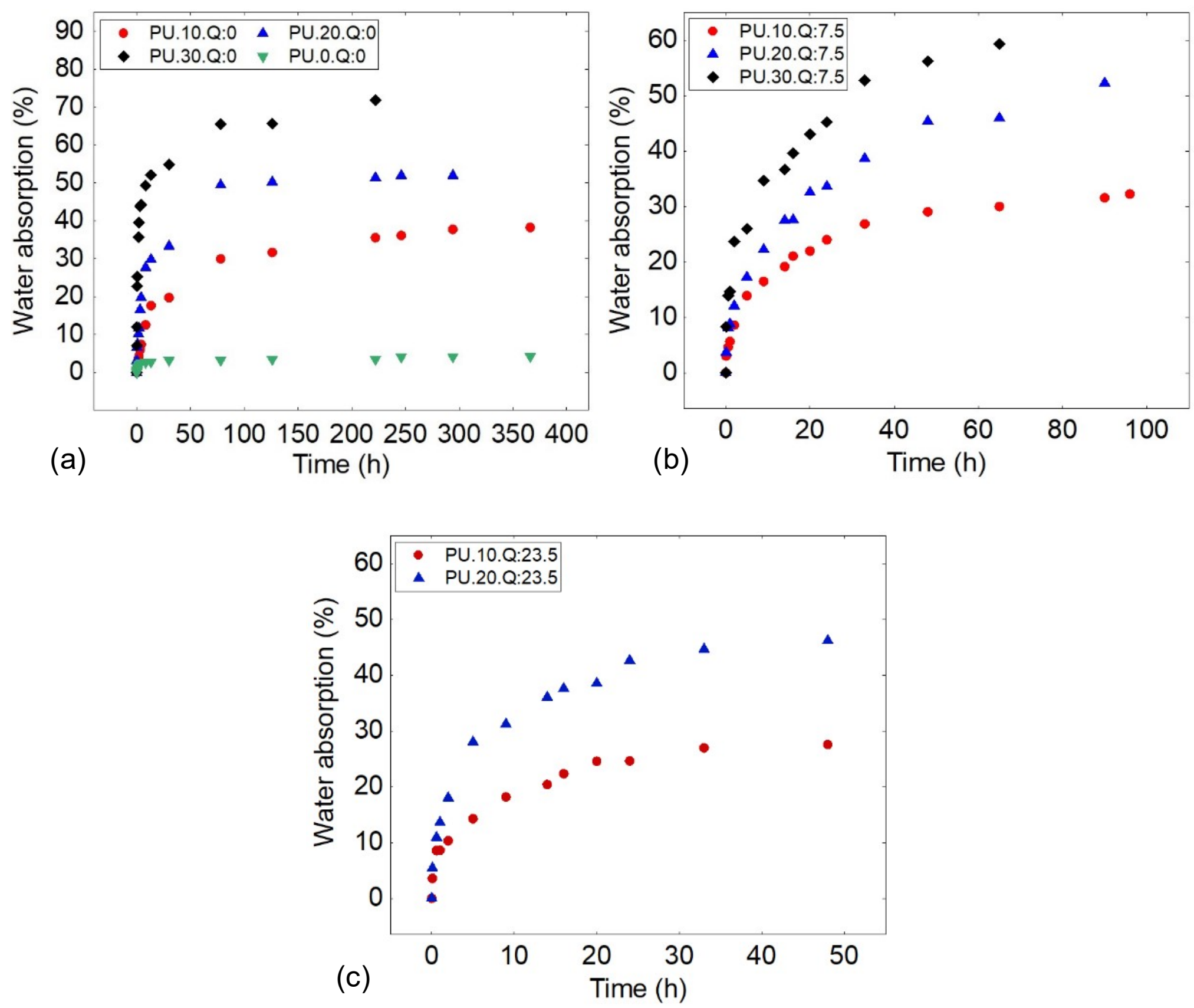

**Figure 3.** Water absorption percentage in accordance to the time of the incubation for different percentages of the drug at (a) static state, (b) the flow rate of 7.5 ml/s, and (c) 23.5 ml/s. (In PU.x.Q:y: x is indicator of drug percentage and y is indicator of flow rate (ml/s))

### *3.3. Cumulative drug release at diverse flow rates and drug percentages*

Figure 4 (a) and (b) show the drug release percentage for the three loaded PU cases in the case of static and dynamic with the flow rate of 7.5 ml/s.

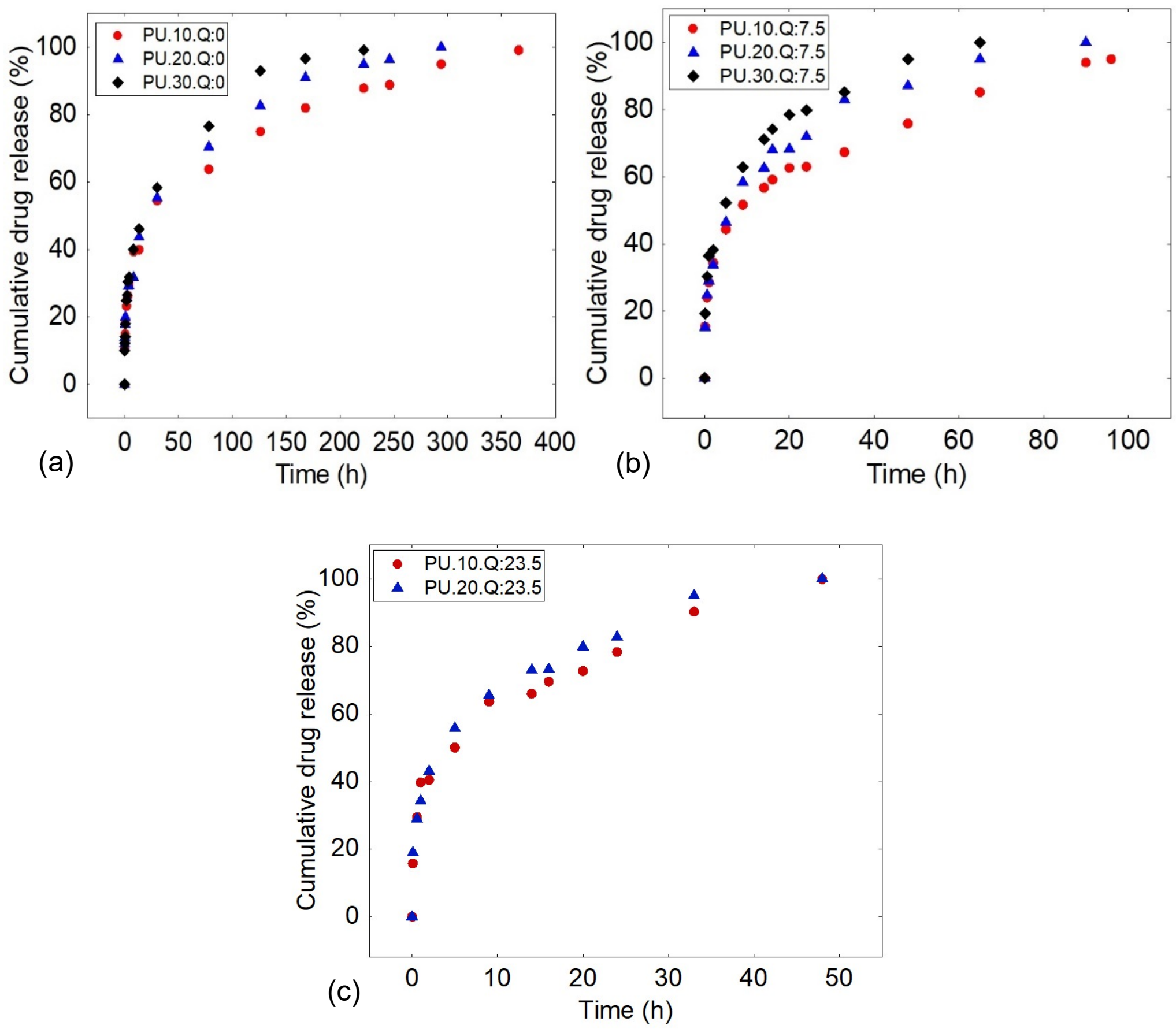

**Figure 4:** Cumulative drug release in accordance with the time of the incubation for different percentages of the drug and at (a) static state, (b) the flow rates of 7.5 ml/s, and (c) 23.5 ml/s

Results are given in figure 4 (a) and (b), depict the drug release at 10, 20, and 30 percentage of initial dosage under static and dynamic conditions, respectively. It is clear that the release time for the samples with the higher drug dosage has decreased. On the other hand, by comparing figure 4 (a), (b), and (c) one can note that with increasing flow rate, drug release is occurring at a higher rate. However, these results need more analysis, which is discussed in section 5.

### *3.4. Understanding of drug release mechanisms: strategy to investigate*

Mathematical models have always been one of the most effective ways to improve the design of different carriers for the drug delivery systems. In addition, it has always been important to determine the mechanisms of drug release and release kinetics for various systems [31-34]. For this purpose, in this section, we will investigate the steps of drug release from the carrier as well as the precise mechanisms of drug release from the carrier using fitting experimental data with different physical models [7, 10, 14, 35, 36]. Additionally, all the regressions were performed with 95% confidence intervals.

In order to investigate the mechanisms, first finding the steps of the release profile is considered. The experimental data was investigated using Higuchi model, which has been used extensively in several studies to evaluate the stages of drug release from carries [27, 37-39]. The Higuchi equation is as follows [40]:

$$Q = \sqrt{D(2C - C_s)C_s t} \qquad \text{Eq. 3}$$

where Q is the amount of drug released at time t per unit area, C is the amount of primary drug in the matrix per unit volume, $C_s$ is the drug solubility in the matrix media and D is the diffusion coefficient of the drug molecules in the carrier. In general, the simplified Higuchi model is defined as below [40]:

$$\frac{M_t}{M_\infty} = K_H\sqrt{t} \qquad \text{Eq. 4}$$

where $M_t$ and $M_\infty$, are drug released at t and infinitive time, respectively, and $K_H$ is a kinetic constant.

After identifying the steps, Korsmeyer-Peppas was applied to each curve in order to detect the mechanisms whether it is diffusion or degradation and swelling controlled. For this purpose, the Peppas model was first evaluated [41]:

$$\frac{M_t}{M_\infty} = kt^n \qquad \text{Eq. 5}$$

where k is a constant and the exponent n determines the mechanism of drug release from the carrier.

As per to the coefficient of categorization for a thin film, when $n < 0.5$, the semi-Fickian mechanism is presented while the $n = 0.5$ the drug release mechanism can be controlled by Fickian diffusion. Whereas $0.5 < n < 1$ the mechanism can be followed by anomalous transport (diffusion and degradation: non-Fickian). Once $n = 1$ the release can be controlled by degradation.

To further investigate the mechanism of drug release from carriers; Zero-order, First-order, and Peppas-Sahlin models were examined. Zero-order model is utilized where drug release is controlled by matrix degradation [42]:

$$\frac{M_t}{M_\infty} = kt \qquad \text{Eq. 6}$$

Here, model order is the order of release rate and not the order of cumulative release. The first-order release in the case where drug release experiences an exponential behavior, also, the dissolution and geometry of the matrix are considered in this model [42]:

$$\frac{M_t}{M_\infty} = 1 - \exp(-kt) \quad \text{Eq. 7}$$

In Peppas-Sahlin equation, drug release is controlled by the diffusion and relaxation mechanism [43]:

$$\frac{M_t}{M_\infty} = k_1 t^m + k_2 t^{2m} \quad \text{Eq. 8}$$

where $k_1$ and $k_2$ are kinetic constants and m is the diffusion exponent.

Moreover, Peppas-Sahlin model is used to calculate the percentage of the contribution of the mechanism of the diffusion during the release. In Peppas-Sahlin equation, by considering the following relation [43]:

$$F = \frac{1}{1+\frac{k_2}{k_1} t^m} \quad \text{Eq. 9}$$

where $k_1$ and $k_2$ are constants obtained from equation 6, the contribution of the diffusion mechanism at different times can be investigated. Figure 5 shows the algorithm for following this strategy.

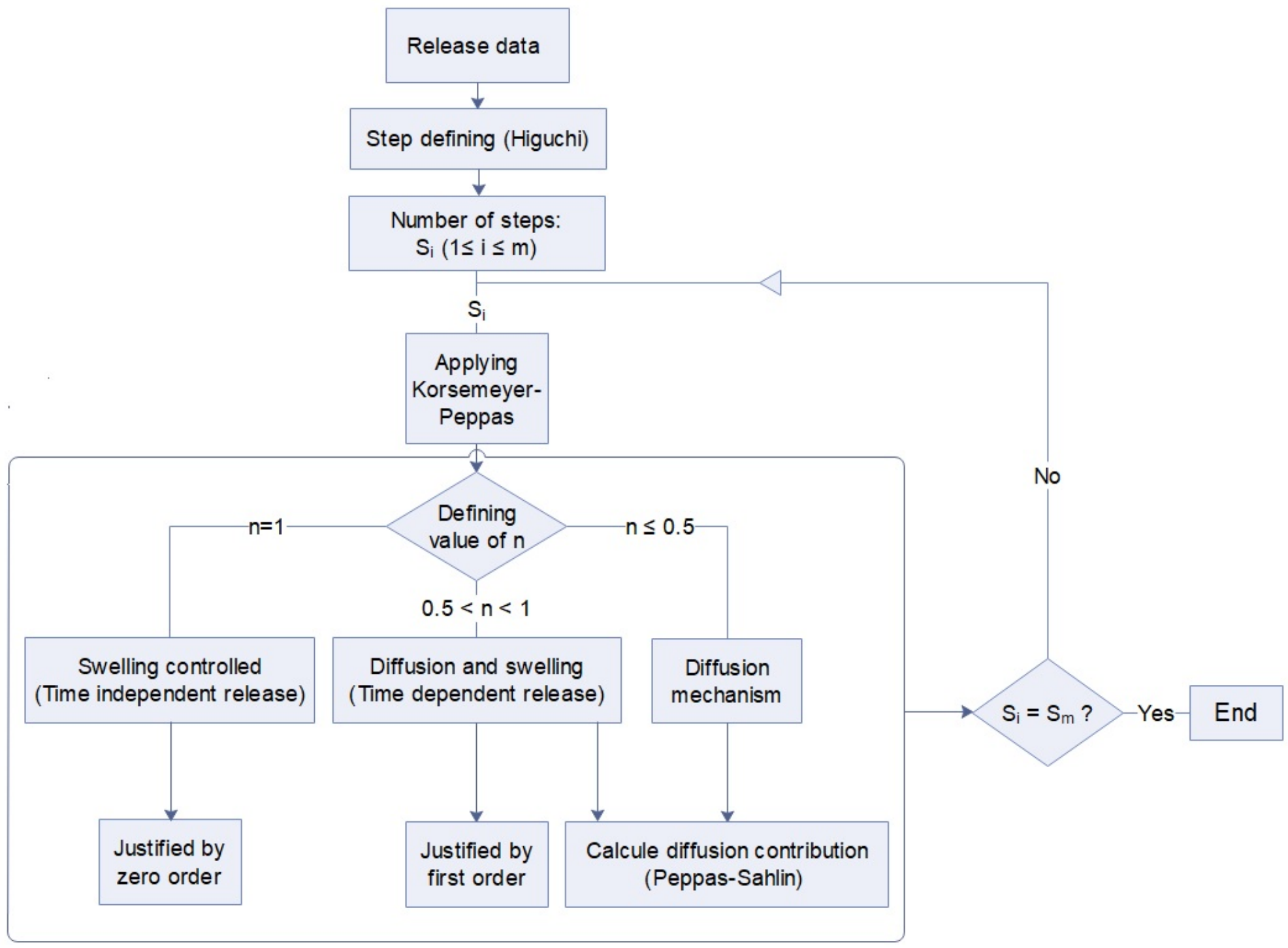

**Figure 5.** Algorithm for defining the mechanisms contributed during the release from a drug loaded carrier

### *3.5. Identification of drug release steps: Higuchi equation*

The experimental results have been analyzed by Higuchi equation (figure 6 and table 1). Fitting the experimental data at the static and dynamic conditions with different dosages by Higuchi model, presented two-step drug release from the PU films.

The regression results of the Higuchi model for each step are presented in table 1. The correlation coefficients for all the samples were high and near to 0.99. In general, comparing the values of the $K_H$ from table 1, show that the kinetic constants in the first and second steps are increased by flow rate.

**Table 1.** Values related to the Higuchi model by fitting the experimental results

| Test condition | $k_H$ | | $R^2$ | |
|---|---|---|---|---|
| | Step 1 | Step 2 | Step 1 | Step 2 |
| **PU.10.Q:0** | 0.108 | 0.037 | 0.98 | 0.99 |
| **PU.20.Q:0** | 0.108 | 0.049 | 0.98 | 0.99 |
| **PU.30.Q:0** | 0.108 | 0.064 | 0.99 | 0.98 |
| **PU.10.Q:7.5** | 0.175 | 0.071 | 0.98 | 0.99 |
| **PU.20.Q:7.5** | 0.175 | 0.071 | 0.98 | 0.98 |
| **PU.30.Q:7.5** | 0.175 | 0.071 | 0.97 | 0.98 |
| **PU.10.Q:23.5** | 0.285 | 0.106 | 0.98 | 0.99 |
| **PU.20.Q:23.5** | 0.285 | 0.106 | 0.98 | 0.99 |

From these curves, the below conclusion can be drawn:

- The number of steps seems to be independent of the considered parameters.
- Second step gives another kinetic of the release more than the first step. Therefore, it notes the presence of the other mechanism(s) or it can be the same mechanism as first step but with different kinetics due to some reasons.
- The release kinetics of the first step is independent of different drug percentages for different flow rates.
- One can define that the threshold time, which is the time where the kinetic changes, are different for dynamic flow rates; however, it is the same for the static state shown in figure 7(a).
- In static state, the rate of drug release in the second step increases by increasing the drug dosage. However, it is constant for dynamic flow rates (figure 7(b)). On the other hand, the rate of release is more influenced by the drug dosage in the static state.

- For dynamic flow rates, it is notable that by increasing the drug percentage the threshold time is increased (figure 7(a)). The later confirms that when the drug content is increased at dynamic state more particles can participate in the release at first step. This means there is a delay time in kinetic reduction by increasing the drug dosage perhaps due to the presence of flow or agitation.

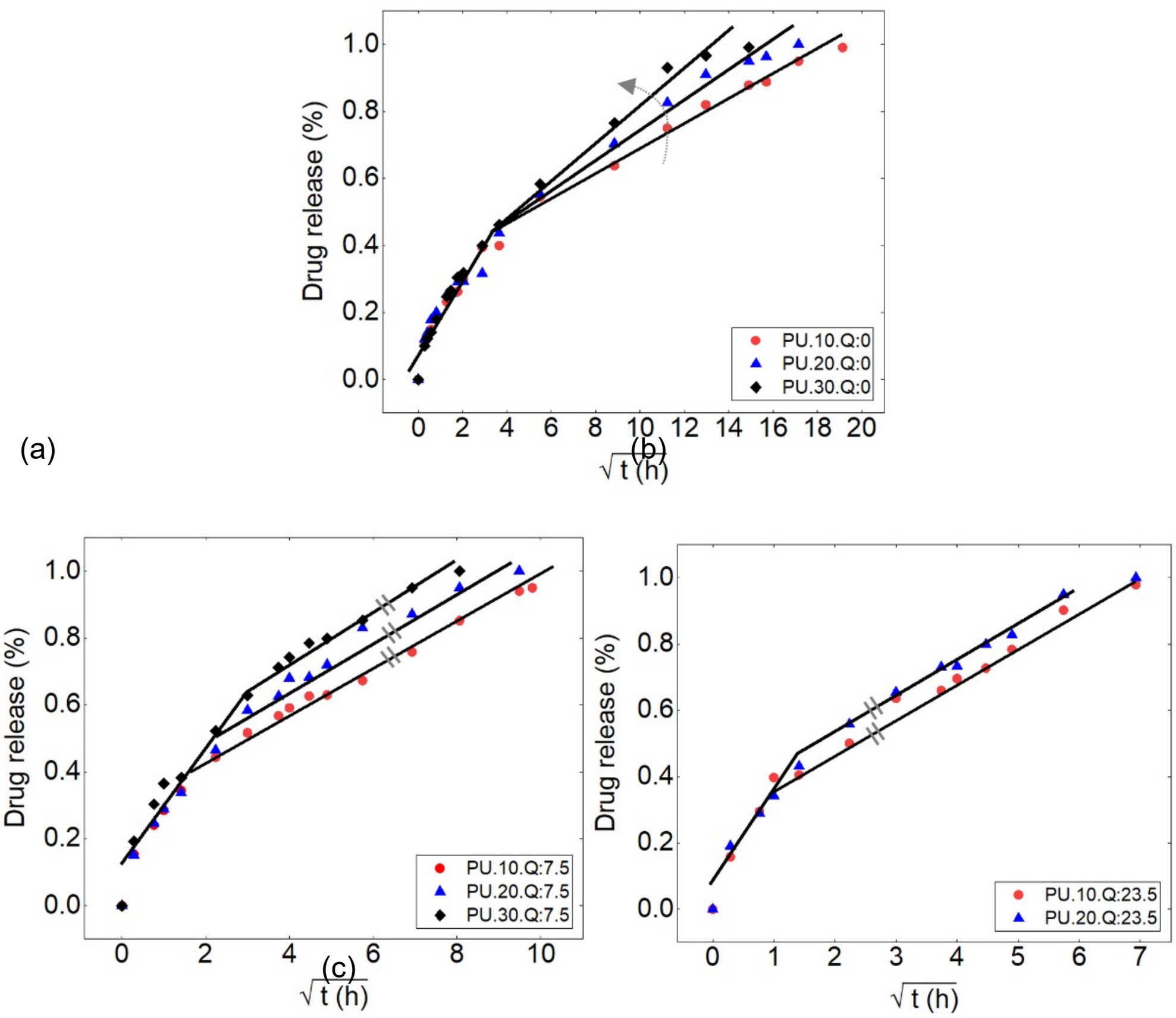

**Figure 6.** Step analyzing by Higuchi model for the experimental tests of drug release with different drug loads (10, 20, and 30%) at flow rates of (a) 0, (b) 7.5 ml/s, and (c) 23.5 ml/s

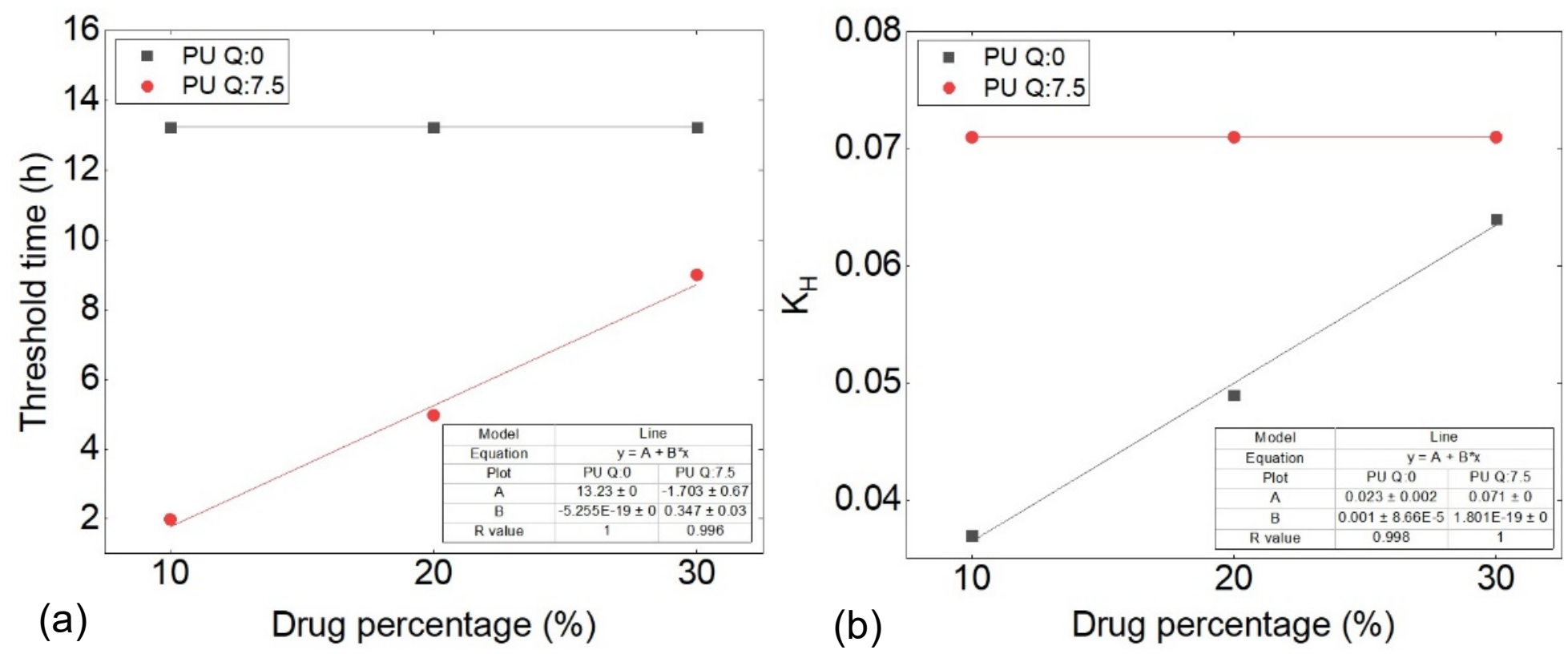

**Figure 7.** Comparing (a) the threshold times and (b) $K_H$ (release kinetic) for the second step of drug release; at different drug percentage for two flow rates of zero and 7.5 ml/s

### *3.6. Investigation of the drug release mechanism*

In the previous section, dual-stage drug release from films was observed. In this section, the mechanisms of drug release under static and dynamic conditions were investigated.

#### 3.6.1. Static state

In this section, the mechanism of drug release at the different dosages of Diclofenac from PU.10.Q:0, PU.20.Q:0 and PU.30.Q:0 is studied. To investigate and intuition the mechanisms of drug release from carriers, equations 5-7 were fitted. The fittings were applied to each step; the results of these models are presented in table 2. Figure 8 shows the curves related to the fitting of the Korsemeyer- Peppas model.

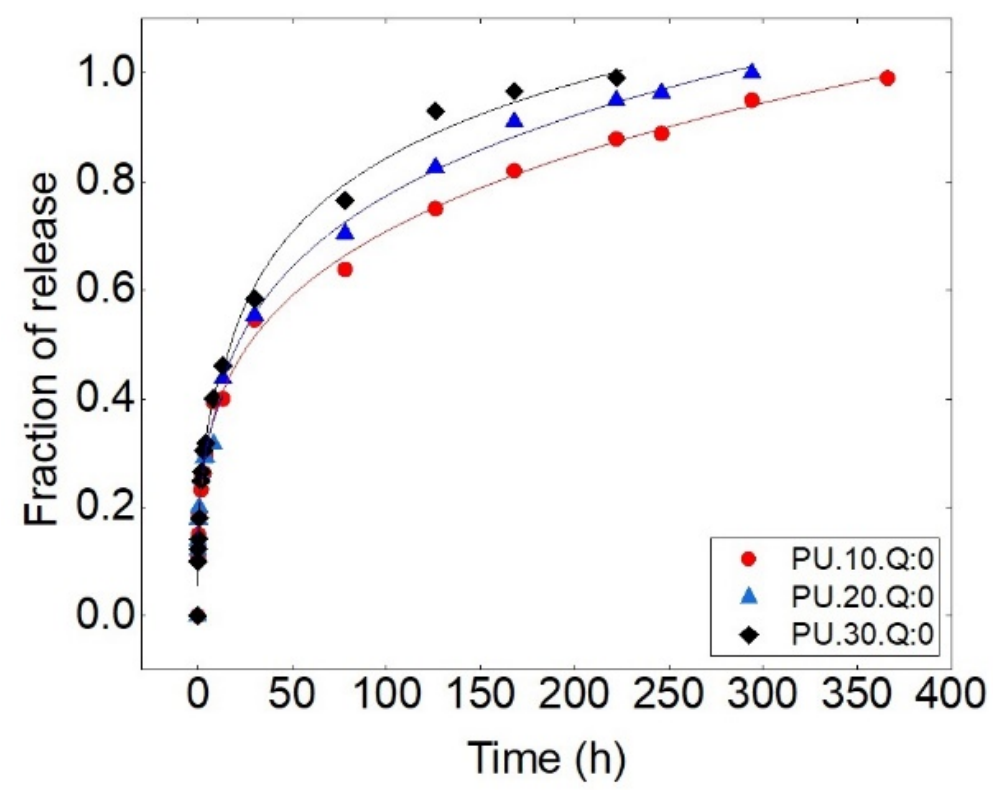

**Figure 8.** Regression results of PU.10.Q:0, PU.20.Q:0 and PU.30.Q:0 in two steps with Korsemeyer- Peppas model

The constants related to the fitting of equations 5, 6, and 7 related to Korsemeyer-Peppas, zero-order, and first-order models, respectively, are presented in table 2. The unfavorable fit of equations 6 and 7 indicates the ignorable of polymer degradation and non-integration of the dissolution, respectively. The regression results of the Korsemeyer- Peppas model presented in table 2 show the correlation coefficients for all three samples were higher than 0.98. The n values are less than 0.5. It is noteworthy that, the amount of n value less than 0.5 indicates the pseudo-Fickian mechanism [44, 45] for drug release from the loaded PU. Therefore, in the case of the static state at two steps for three different percentages diffusion is the involved mechanism but it is not the only mechanism intervene.

**Table 2.** Regression constants for different mathematical models for PU.10, 20, 30.Q:0

| Polymeric films | Korsemeyer- Peppas | | | | | | Zero-order | | | | First order | | | |
|---|---|---|---|---|---|---|---|---|---|---|---|---|---|---|
| | K | | n | | $R^2$ | | K | | $R^2$ | | K | | $R^2$ | |
| | Step | Step | Step | Step | Step | Step | Step 1 | Step | Step | Step | Step | Step | Step | Step |

| | 1 | 2 | 1 | 2 | 1 | 2 | | 2 | 1 | 2 | 1 | 2 | 1 | 2 |
|---|---|---|---|---|---|---|---|---|---|---|---|---|---|---|
| PU.10.Q:0 | 0.19 | 0.23 | 0.29 | 0.25 | 0.99 | 0.99 | 0.042 | 0.002 | 0.12 | 0.95 | 0.063 | 0.015 | 0.29 | 0.78 |
| PU.20.Q:0 | 0.16 | 0.51 | 0.30 | 0.18 | 0.98 | 0.99 | 0.042 | 0.002 | 0.12 | 0.95 | 0.062 | 0.022 | 0.25 | 0.87 |
| PU.30.Q:0 | 0.21 | 0.60 | 0.31 | 0.16 | 0.99 | 0.98 | 0.046 | 0.002 | 0.19 | 0.91 | 0.072 | 0.029 | 0.57 | 0.93 |

### 3.6.2. Dynamic state

For identifying the mechanisms in dynamic state, the flow rate of 7.5 ml/s was chosen. The results of regression by equations 5, 6 and 7 on the data are presented in table 3, and as per the obtained value, it can be stated that the mechanism of drug release represents pseudo-diffusion. Like as the static one here also the curves don't show good fit to the equations 6, 7; therefore, it indicates that the release is unintegrated into the dissolution or degradation of the matrix. The semi-Fickian mechanism is controlling the drug release at all two steps. However, for considering the degradation in the test situation degradation test were conducted in the continuous flow state where, the three pure polyurethane samples were immersed in the aquatic environment for 28 days. The results demonstrated no mass loss during this test. As the mass loss is an indicator of the degradation the material is facing the first type of degradation where it experiences a decrease in mechanical properties but no weight loss during the test period [40, 41]. Therefore, referring to the experimental degradation test of polyurethane samples, no degradation was observed. Figure 9 shows the curves related to the fitting of Korsemeyer- Peppas model applied on these results.

**Table 3.** Regression constants for different mathematical models for PU.10, 20, 30.Q:7.5

| Polymeric films | Korsemeyer- Peppas | | | | | | Zero-order | | | | First order | | | |
|---|---|---|---|---|---|---|---|---|---|---|---|---|---|---|
| | K | | n | | $R^2$ | | K | | $R^2$ | | K | | $R^2$ | |
| | Step 1 | Step 2 | Step 1 | Step 2 | Step 1 | Step 2 | Step 1 | Step 2 | Step 1 | Step 2 | Step 1 | Step 2 | Step 1 | Step 2 |
| PU.10.Q:7.5 | 0.28 | 0.08 | 0.27 | 0.46 | 1 | 1 | 0.11 | 0.005 | 0.28 | 0.98 | 0.17 | 0.05 | 0.59 | 0.71 |
| PU.20.Q:7.5 | 0.29 | 1.11 | 0.28 | 0.12 | 1 | 0.99 | 0.113 | 0.006 | 0.40 | 0.93 | 0.18 | 0.07 | 0.64 | 0.84 |
| PU.30.Q:7.5 | 0.35 | 4.16 | 0.24 | 0.04 | 0.99 | 1 | 0.13 | 0.007 | 0.11 | 0.93 | 0.22 | 0.88 | 0.71 | 0.84 |

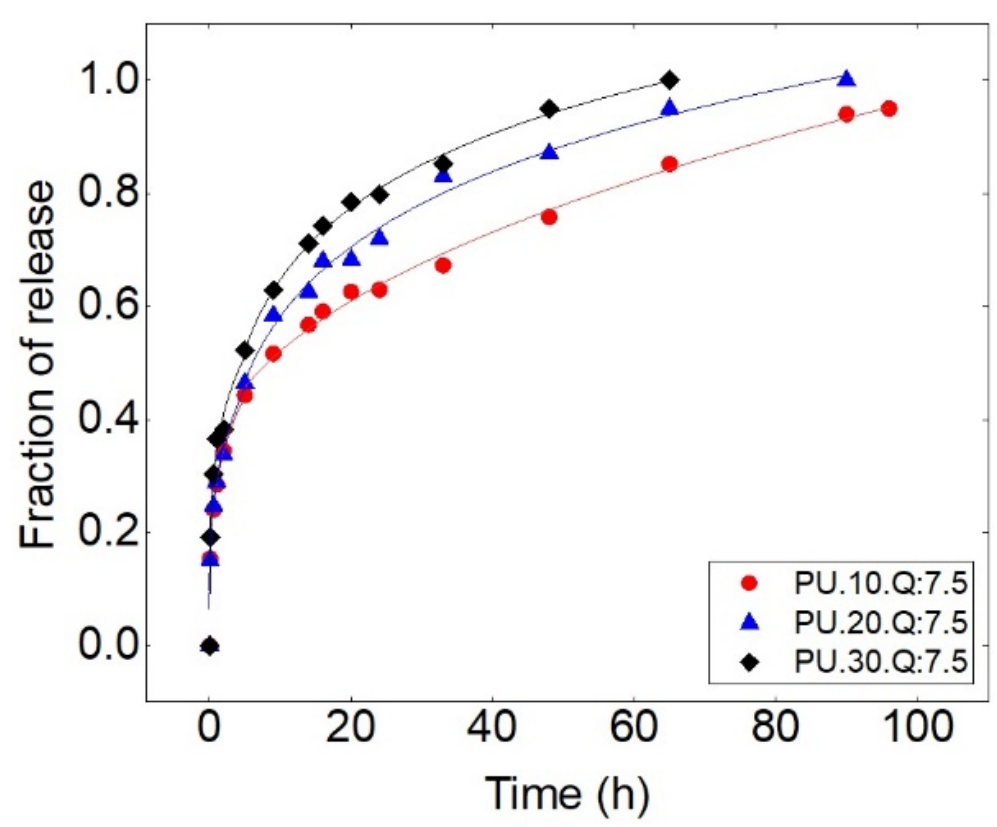

**Figure 9.** Regression results of PU.10, 20, 30.Q:7.5 with Korsemeyer- Peppas model

### *3.7. Contribution of the diffusion mechanism*

#### 3.7.1. Effect of flow rates

After confirming the diffusion transport of the drug from the carriers, the contribution of diffusion mechanisms for the samples with 10% of drug at the flow rates of 0, 7.5, 23.5 ml/s by equations 8 and 9 was calculated. Figure 10 shows the contribution of the diffusion mechanism against drug release times for PU.10.Q:0, PU.10.Q:7.5, and PU.10.Q:23.5. It shows that the portion of the diffusion mechanism decreased along with increasing the time for all three samples and also it is clear that an increase of flow rate from 0, 7.5 ml/s to 23.5 ml/s, causes a decrease in the portion of the diffusion mechanism in drug release.

The decrease in the contribution of the diffusion mechanism over time will be justified by the decrease in the concentration gradient by the time because the diffusion mechanism is controlled by the potential chemical gradient. As mentioned, drug release can involve a variety of mechanisms, common mechanism for drug delivery based on polyurethanes can be diffusion [37]. In the previous section, the contribution of the diffusion mechanism was obtained by equation 9 and it was shown that the diffusion mechanism was the dominant mechanism throughout the drug release period, but according to figure ١٠, all the drugs were not released just by this mechanism. It is also noteworthy that if the mechanism of drug release was uniquely based on diffusion, the n value obtained from equation 5 should be equal to the m value obtained from equation 8 [43] which in accordance with the calculations these values were not equal.

Therefore, the contribution of the other mechanisms takes attention. The contribution of the other mechanisms can be calculated by the following relation:

Total drug release = (diffusion mechanism + other mechanisms) contributions to release

The results of the degradation of polyurethane at different times showed no degradation. Therefore, drug release cannot be attributed to the carrier's degradation or dissolution.

In the first step for the matrix carrier especially with the undissolved drug particles, the first moments of the release are probably related to the phenomena of burst release. As the studies of the burst-release have shown this mechanism inevitably occurs in the first liberation period and will continue until the release of the drug is stable [46, 47]. Referring to the initial values of the burst release obtained from the figure 10 indicate that the burst release is increased by changing the state of the flow from static to continuous, whereas it shows fewer differences for two cases of flow rates.

For the second step, another mechanism in which the drug can be released is based on osmotic pump, this mechanism can be created by osmotic pressure and it is not based on diffusion. One of the reasons for this mechanism is the absorption of water into the polymer. Also, the polymer degradation must be negligible and the channels of the polymer should be larger than 60 microns [48-50]. This value is justified by the microscopic images (figure 11). In this figure the dark round holes refer to the cavities, where the lighter holes refer to the bubbles created during the preparation which may contain the drug inside of them.

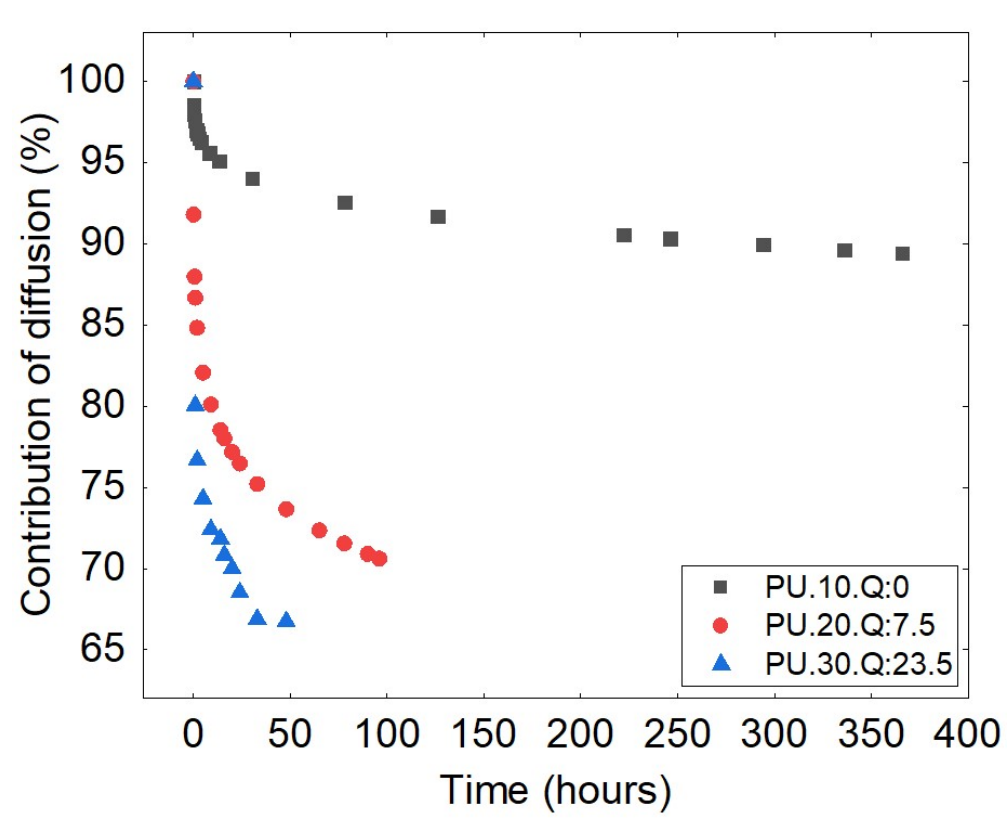

**Figure 10.** Contribution of diffusion in accordance with the time of the release at different flow rates for PU+10% drug

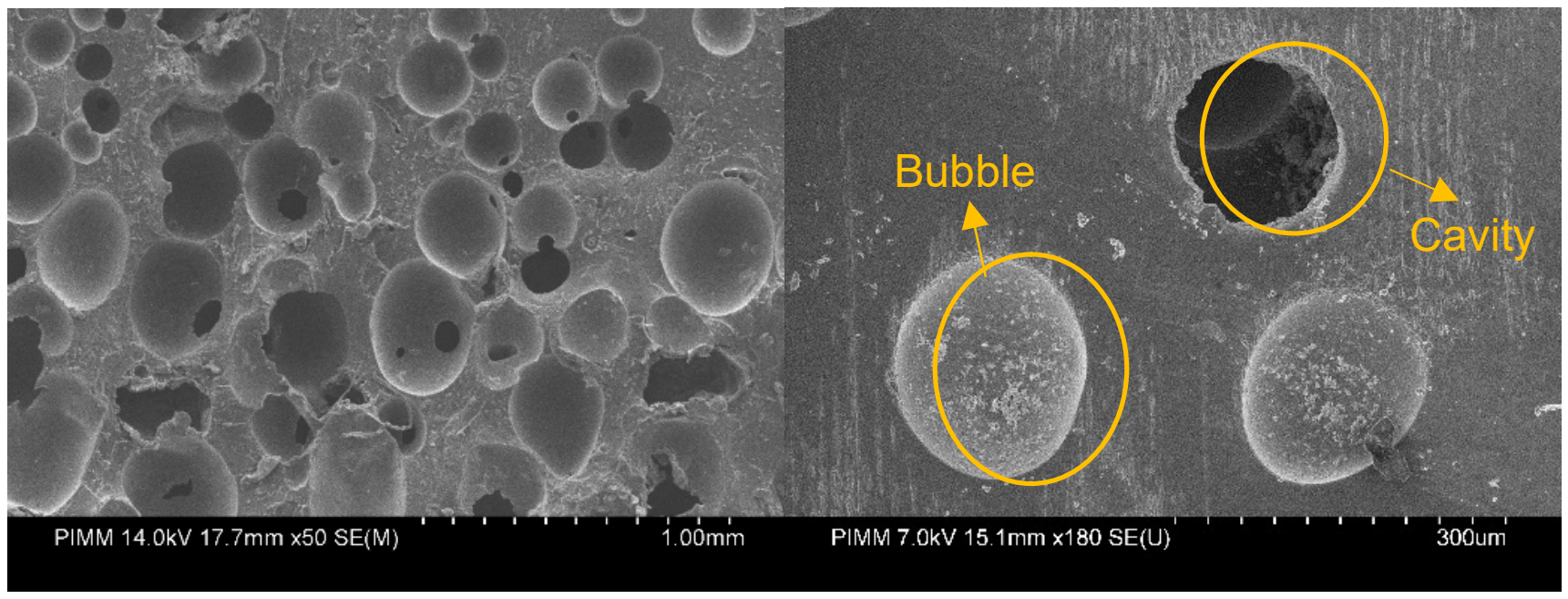

**Figure 11.** Microscopic images representing the size of hole (PU.10. Q:7.5 after 1 hour the test)

The results of water absorption showed the amount of water absorbed by PU.10.Q:0, PU.10.Q:7.5 and PU.10.Q:23.5 increased with time. Moreover, increasing the flow rate increases water penetration into the polymer channels and pores. On the other hand, the osmotic pumping is based on the osmotic pressure resulting from water absorption and the solutes dissolved in it [47, 50]. Depending on the type of the matrix as it is semipermeable or non-permeable with the solutes inside, the osmotic pressure can result in the swelling or shrinkage of the samples, respectively. According to the semi-permeable samples used in this study swelling is the probable case. However, as it is mentioned in the part 3.2 the swelling in this case is also Fickian-controlled. Thus, it can be said that increasing the rate of water uptake over time and increasing the flow rate causes the release of the drug through osmotic pressure [48, 50]. Consequently, the contribution of the diffusion mechanism to the drug release is decreased by increasing the flow rate and the time of the incubation.

### 3.7.2. Different Drug percentages

Figure 12 shows the contribution of diffusion mechanisms for the samples with different drug percentages at the flow rate of 7.5 ml/s by equations 8 and 9. As shown in figure 12, the contribution of the diffusion mechanism decreases over time for all three cases. Furthermore, the non-equilibrium coefficients n and m obtained from equations Korsemeyer- Peppas and Peppas-Sahlin showed that drug release was not affected only by diffusion. In the previous section, it was shown that the amount of water absorbed by the carrier increased over time. On the other hand, with increasing drug content in polymer films, water absorption increased which reduced the influence of the diffusion mechanism on drug release and increased the contribution of the other mechanisms.

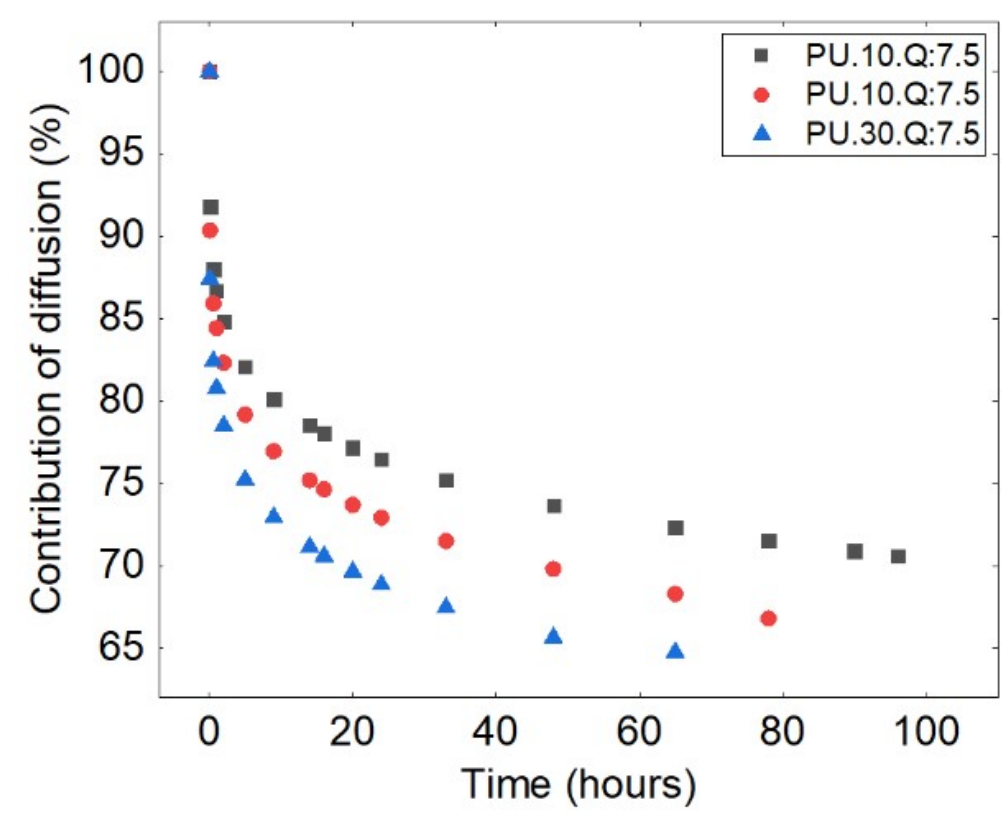

**Figure 12.** Contribution of diffusion in accordance with the time of the release at different initial drug percentages at the Q=7.5 ml/s

Referring to the figure 12 at the first step as the contribution of the diffusion is decreased at the starting point of release for the samples with 10, 20 and 30 percentages of drug. Therefore, release related to the contribution of other phenomena at the first period of the release respectively for 10, 20 and 30 percentages of drug increases. As mentioned above for the matrix sample at the first moments of the release the probable phenomenon is the burst release which is less related to the water absorption. As the percentage of the drug increases the probability of the drug particles to stay on the surface of the polymer matrix increases. It is therefore by increasing the drug percentage the value related to the burst release increases.

In the second step, the other mechanism which is contributed is the osmotic pressure [49, 50] which is increased by the water absorption and increasing the free space by releasing the more drugs. It is notable more the drug percentage, more the contribution of the osmotic pressure. The system is more non-equilibrium where the drug percentage increases, therefore activation energy for contributing osmotic pump increases. Moreover, hydrophilicity of the sample due to the more hydrophilic drug increases, hence water absorption increases, and vapor pressure would be another reason for commencing the osmotic pressure. Additionally, another parameter affecting the osmotic pressure is the permeability of the matrix, where increasing the drug content increases the permeability in the matrix. Therefore, the release of the drug through osmotic pressure will depend on the amount of drug loaded and consequently water absorbed into the polymer.

## 4. Conclusion

In this work, the release of diclofenac from a polyurethane film loaded with various drug percentages was studied under static and dynamic flow conditions. Water absorption and cumulative drug release values were identified for all flow rates and drug dosages.

The experimental data was investigated using Higuchi model and proposed calculation algorithm to evaluate the stages of drug release from carries and contributed mechanisms. Drug release took place in two stages for static and dynamic states which were independent of the considered parameters. The second step gives another kinetic of the release more than the first step. Therefore, it notes the presence of the other mechanism(s), or it can be the same mechanism as the first step but with different kinetics due to some reasons. One can notice that the release kinetics of the first step was independent of different drug percentages for different flow rates. Threshold time, which is the time where the kinetic changes, are different for dynamic flow rates; however, it is the same for the static. In addition, the rate of release is more influenced by the drug dosage in the static state. There was a delay time in kinetic reduction by increasing the drug dosage perhaps due to presence of flow or agitation.

The results allowed the identification of three drug delivery of mechanisms; burst-release, diffusion and osmotic pressure mechanisms. The diffusion represents the dominant mechanism in all periods of delivery. However, the contribution of the burst-release throughout the initial time and the osmotic pressure during the second step is accompanied by the diffusion. The proportion of drugs delivered in accordance with time for each of these mechanisms changes during the release period. In addition, the contribution of the other mechanisms apart from diffusion increases with the flow rate and as the percentage of drugs.

## 5. Conflicts of Interest

The authors declare that they have no conflicts of interest.

## 6. References

[1] G. Tiwari *et al.*, "Drug delivery systems: An updated review," vol. 2, no. 1, p. 2, 2012.

[2] W. B. Liechty, D. R. Kryscio, B. V. Slaughter, N. A. J. A. r. o. c. Peppas, and b. engineering, "Polymers for drug delivery systems," vol. 1, pp. 149-173, 2010.

[3] T. K. Dash and V. B. Konkimalla, "Poly-є-caprolactone based formulations for drug delivery and tissue engineering: A review," *Journal of Controlled Release,* vol. 158, no. 1, pp. 15-33, 2012.

[4] T. J. Johnson, K. M. Gupta, J. Fabian, T. H. Albright, and P. F. Kiser, "Segmented polyurethane intravaginal rings for the sustained combined delivery of antiretroviral agents dapivirine and tenofovir," *European Journal of Pharmaceutical Sciences,* vol. 39, no. 4, pp. 203-212, 2010.

[5] F. R. Nezami, L. S. Athanasiou, and E. R. Edelman, "Endovascular drug-delivery and drug-elution systems," in *Biomechanics of Coronary Atherosclerotic Plaque*: Elsevier, 2020, pp. 611-652.

[6] C. Bode, H. Kranz, A. Fivez, F. Siepmann, and J. Siepmann, "Often neglected: PLGA/PLA swelling orchestrates drug release: HME implants," *Journal of Controlled Release,* vol. 306, pp. 97-107, 2019.

[7] Y. Fu and W. J. Kao, "Drug release kinetics and transport mechanisms of non-degradable and degradable polymeric delivery systems," *Expert opinion on drug delivery,* vol. 7, no. 4, pp. 429-444, 2010.

[8] Q. Guo, P. T. Knight, and P. T. Mather, "Tailored drug release from biodegradable stent coatings based on hybrid polyurethanes," *Journal of controlled release,* vol. 137, no. 3, pp. 224-233, 2009.

[9] J. Y. Cherng, T. Y. Hou, M. F. Shih, H. Talsma, and W. E. Hennink, "Polyurethane-based drug delivery systems," *International journal of pharmaceutics,* vol. 450, no. 1-2, pp. 145-162, 2013.

[10] D. Y. Arifin, L. Y. Lee, and C.-H. Wang, "Mathematical modeling and simulation of drug release from microspheres: Implications to drug delivery systems," *Advanced drug delivery reviews,* vol. 58, no. 12-13, pp. 1274-1325, 2006.

[11] M. D. Campiñez *et al.*, "Development and characterization of new functionalized polyurethanes for sustained and site-specific drug release in the gastrointestinal tract," *European journal of pharmaceutical sciences,* vol. 100, pp. 285-295, 2017.

[12] A. Basu, S. Farah, K. Kunduru, S. Doppalapudi, W. Khan, and A. Domb, "Polyurethanes for controlled drug delivery," in *Advances in Polyurethane Biomaterials*: Elsevier, 2016, pp. 217-246.

[13] M. B. Lowinger, S. E. Barrett, F. Zhang, and R. O. Williams, "Sustained release drug delivery applications of polyurethanes," *Pharmaceutics,* vol. 10, no. 2, p. 55, 2018.

[14] S. A. Fouad, E. B. Basalious, M. A. El-Nabarawi, and S. A. Tayel, "Microemulsion and poloxamer microemulsion-based gel for sustained transdermal delivery of diclofenac epolamine using in-skin drug depot: in vitro/in vivo evaluation," *International journal of pharmaceutics,* vol. 453, no. 2, pp. 569-578, 2013.

[15] N. Abbasnezhad, M. Shirinbayan, A. Tcharkhtchi, and F. Bakir, "In vitro study of drug release from various loaded polyurethane samples and subjected to different non-pulsed flow rates," *Journal of Drug Delivery Science and Technology,* p. 101500, 2020.

[16] T. T. N. Huynh *et al.*, "Characterization of a polyurethane-based controlled release system for local delivery of chlorhexidine diacetate," *European Journal of Pharmaceutics and Biopharmaceutics,* vol. 74, no. 2, pp. 255-264, 2010.

[17] H. Patel, D. R. Panchal, U. Patel, T. Brahmbhatt, and M. Suthar, "Matrix type drug delivery system: A review," *J Pharm Sci Biosci Res,* vol. 1, no. 3, pp. 143-51, 2011.

[18] W.-W. Yang and E. Pierstorff, "Reservoir-based polymer drug delivery systems," *Journal of laboratory automation,* vol. 17, no. 1, pp. 50-58, 2012.

[19] A. Raval, J. Parikh, and C. Engineer, "Mechanism of controlled release kinetics from medical devices," *Brazilian Journal of Chemical Engineering,* vol. 27, no. 2, pp. 211-225, 2010.

[20] L. Lamoudi, J. C. Chaumeil, and K. Daoud, "Swelling, erosion and drug release characteristics of sodium diclofenac from heterogeneous matrix tablets," *Journal of Drug Delivery Science and Technology,* vol. 31, pp. 93-100, 2016.

[21] D. Ailincai, G. Gavril, and L. Marin, "Polyvinyl alcohol boric acid–A promising tool for the development of sustained release drug delivery systems," *Materials Science and Engineering: C,* vol. 107, p. 110316, 2020.

[22] B. Petersen and S. Rovati, "Diclofenac epolamine (Flector®) patch," *Clinical drug investigation,* vol. 29, no. 1, pp. 1-9, 2009.

[23] S. McGinty and G. Pontrelli, "A general model of coupled drug release and tissue absorption for drug delivery devices," *Journal of controlled release,* vol. 217, pp. 327-336, 2015.

[24] H. Gasmi *et al.*, "Towards a better understanding of the different release phases from PLGA microparticles: Dexamethasone-loaded systems," *International journal of pharmaceutics,* vol. 514, no. 1, pp. 189-199, 2016.

[25] F. Farahmandghavi, M. Imani, and F. Hajiesmaeelian, "Silicone matrices loaded with levonorgestrel particles: impact of the particle size on drug release," *Journal of Drug Delivery Science and Technology,* vol. 49, pp. 132-142, 2019.

[26] K. Škrlová *et al.*, "Biocompatible polymer materials with antimicrobial properties for preparation of stents," *Nanomaterials,* vol. 9, no. 11, p. 1548, 2019.

[27] N. Zirak, A. B. Jahromi, and E. Salahinejad, "Vancomycin release kinetics from Mg–Ca silicate porous microspheres developed for controlled drug delivery," *Ceramics International,* vol. 46, no. 1, pp. 508-512, 2020.

[28] P. Fregolente, H. Gonçalves, M. Wolf Maciel, and L. Fregolente, "Swelling Degree and Diffusion Parameters of Poly (Sodium Acrylate-Co-Acrylamide) Hydrogel for Removal of Water Content from Biodiesel," *Chem. Eng. Trans,* vol. 65, 2018.

[29] B. Taşdelen, N. Kayaman-Apohan, O. Güven, and B. M. Baysal, "Swelling and diffusion studies of poly (N-isopropylacrylamide/itaconic acid) copolymeric hydrogels in water and aqueous solutions of drugs," *Journal of applied polymer science,* vol. 91, no. 2, pp. 911-915, 2004.

[30] B. B. Mandal, S. Kapoor, and S. C. Kundu, "Silk fibroin/polyacrylamide semi-interpenetrating network hydrogels for controlled drug release," *Biomaterials,* vol. 30, no. 14, pp. 2826-2836, 2009.

[31] P. Ilgin, H. Ozay, and O. Ozay, "A new dual stimuli responsive hydrogel: Modeling approaches for the prediction of drug loading and release profile," *European Polymer Journal,* vol. 113, pp. 244-253, 2019.

[32] A. Valério, E. Mancusi, F. Ferreira, S. M. G. U. de Souza, A. A. U. de Souza, and S. Y. G. González, "Biopolymer-hydrophobic drug fibers and the delivery mechanisms

for sustained release applications," *European Polymer Journal,* vol. 112, pp. 400-410, 2019.

[33] R. Mohapatra, D. Ray, A. Swain, T. Pal, and P. Sahoo, "Release study of alfuzosin hydrochloride loaded to novel hydrogel P (HEMA-co-AA)," *Journal of applied polymer science,* vol. 108, no. 1, pp. 380-386, 2008.

[34] D. N. Soulas and K. G. Papadokostaki, "Experimental investigation of the release mechanism of proxyphylline from silicone rubber matrices," *Journal of Applied Polymer Science,* vol. 120, no. 2, pp. 821-830, 2011.

[35] C. Algieri, E. Drioli, and L. Donato, "Development of mixed matrix membranes for controlled release of ibuprofen," *Journal of applied polymer science,* vol. 128, no. 1, pp. 754-760, 2013.

[36] P. A. Moradmand and H. Khaloozadeh, "An experimental study of modeling and self-tuning regulator design for an electro-hydro servo-system," in *2017 5th International Conference on Control, Instrumentation, and Automation (ICCIA)*, 2017, pp. 126-131: IEEE.

[37] J. M. Unagolla and A. C. Jayasuriya, "Drug transport mechanisms and in vitro release kinetics of vancomycin encapsulated chitosan-alginate polyelectrolyte microparticles as a controlled drug delivery system," *European Journal of Pharmaceutical Sciences,* vol. 114, pp. 199-209, 2018.

[38] B. G. Trewyn, S. Giri, I. I. Slowing, and V. S.-Y. Lin, "Mesoporous silica nanoparticle based controlled release, drug delivery, and biosensor systems," *Chemical communications,* no. 31, pp. 3236-3245, 2007.

[39] F. Notario-Pérez, A. Martín-Illana, R. Cazorla-Luna, R. Ruiz-Caro, J. Peña, and M.-D. Veiga, "Improvement of Tenofovir vaginal release from hydrophilic matrices through drug granulation with hydrophobic polymers," *European Journal of Pharmaceutical Sciences,* vol. 117, pp. 204-215, 2018.

[40] T. Higuchi, "Mechanism of sustained-action medication. Theoretical analysis of rate of release of solid drugs dispersed in solid matrices," *Journal of pharmaceutical sciences,* vol. 52, no. 12, pp. 1145-1149, 1963.

[41] P. L. Ritger and N. A. Peppas, "A simple equation for description of solute release I. Fickian and non-fickian release from non-swellable devices in the form of slabs, spheres, cylinders or discs," *Journal of controlled release,* vol. 5, no. 1, pp. 23-36, 1987.

[42] A. L. W. Po, L. Wong, and C. Gilligan, "Characterisation of commercially available theophylline sustained-or controlled-release systems: in-vitro drug release profiles," *International Journal of Pharmaceutics,* vol. 66, no. 1-3, pp. 111-130, 1990.

[43] N. A. Peppas and J. J. Sahlin, "A simple equation for the description of solute release. III. Coupling of diffusion and relaxation," *International journal of pharmaceutics,* vol. 57, no. 2, pp. 169-172, 1989.

[44] R.-D. Pavaloiu, A. Stoica-Guzun, M. Stroescu, S. I. Jinga, and T. Dobre, "Composite films of poly (vinyl alcohol)–chitosan–bacterial cellulose for drug controlled release," *International journal of biological macromolecules,* vol. 68, pp. 117-124, 2014.

[45] N.-N. Li, C.-P. Fu, and L.-M. Zhang, "Using casein and oxidized hyaluronic acid to form biocompatible composite hydrogels for controlled drug release," *Materials Science and Engineering: C,* vol. 36, pp. 287-293, 2014.

[46] X. Huang and C. S. Brazel, "On the importance and mechanisms of burst release in matrix-controlled drug delivery systems," *Journal of controlled release,* vol. 73, no. 2-3, pp. 121-136, 2001.

[47] U. Gbureck, E. Vorndran, and J. E. Barralet, "Modeling vancomycin release kinetics from microporous calcium phosphate ceramics comparing static and dynamic immersion conditions," *Acta biomaterialia,* vol. 4, no. 5, pp. 1480-1486, 2008.

[48] E. L. Cussler and E. L. Cussler, *Diffusion: mass transfer in fluid systems*. Cambridge university press, 2009.

[49] J. Hjärtstam, *Ethyl cellulose membranes used in modified release formulations*. Chalmers University of Technology, 1998.

[50] S. Fredenberg, M. Wahlgren, M. Reslow, and A. Axelsson, "The mechanisms of drug release in poly (lactic-co-glycolic acid)-based drug delivery systems—a review," *International journal of pharmaceutics,* vol. 415, no. 1-2, pp. 34-52, 2011.